# Antihelical Edge States in Photonic Metacrystals

Ke-Yi Zeng, Zi-Xuan Gao, Zhou Zhou, Xiao-Dong Chen*, and Jian-Wen Dong
*School of Physics & State Key Laboratory of Optoelectronic Materials and Technologies, Sun Yat-sen University, Guangzhou 510275, China.*
*Corresponding author: chenxd67@mail.sysu.edu.cn

**Antihelical edge states represent a gapless form of boundary transport in which modes with opposite spin counterpropagate along a given boundary, while modes sharing the same spin propagate in the same direction on opposite boundaries. Here we show that such states arise naturally within a modified Kane-Mele framework realized in an electromagnetic-dual photonic metacrystal with staggered bianisotropy. Reversing the spin-dependent next-nearest-neighbor coupling on one sublattice reorganizes the bulk spectrum into a spin-polarized Dirac semimetal characterized by spin-dependent energy shifts of the Dirac cones rather than a bulk gap. This bulk reorganization generates tilted edge-mode dispersions in zigzag ribbons and gives rise to antihelical edge transport coexisting with bulk states. Full-wave simulations confirm the corresponding bulk and edge behaviors. Extending the analysis to finite geometries, simulations further reveal discrete corner-localized modes. A realistic design based on split-ring metaatoms is proposed to realize the required staggered bianisotropy and spin-dependent energy shifts of the Dirac cones. Furthermore, we show that the tensorial freedom of the bianisotropic constitutive relations enables continuous manipulation of the electromagnetic pseudospin basis while preserving the antihelical boundary transport. These results provide a practical route for engineering unconventional gapless boundary transport in photonic Dirac systems.**

***Introduction. ---*** Topological edge states are commonly discussed in systems with bulk band gaps, where boundary-localized modes are spectrally separated from extended bulk states and protected by topological invariants defined over the Brillouin zone [1, 2]. Representative examples include quantum Hall [3-6] and quantum spin Hall (QSH) phases [7-10], in which nontrivial Chern number, spin-Chern number or $Z_2$ invariants guarantee robust chiral or helical boundary transport [11-13]. Within the conventional bulk-edge correspondence framework, the bulk gap plays a central role by defining the topological invariant and ensuring the robustness of boundary modes against disorder and backscattering [14, 15]. However, boundary transport is not restricted to such gapped settings. Recent studies have shown that unconventional boundary transport can also emerge in gapless systems, where edge states coexist with bulk modes. Two-dimensional graphene with gapless Dirac cones at Brillouin zone boundaries provides a representative platform in this context [16, 17]. Although global topological invariants are ill defined in gapless spectra, boundary-localized states may still appear over certain momentum ranges [18-20]. In graphene-like lattices, such states typically connect Dirac points at inequivalent valleys. When the two Dirac cones are energetically displaced while remaining gapless, their projections onto a given boundary occur at different energies. This energy displacement results in tilted edge-state dispersions and therefore gives rise to boundary transport beyond conventional helical or chiral scenarios. A representative example is the antichiral edge state [21-26], in which edge states on opposite boundaries propagate in the same direction. An analogous phenomenon can occur in time-reversal-invariant systems, where spin-up and spin-down states form Kramers pairs, giving rise to antihelical edge states [27-30]. In this case, states with the same spins propagate in the same direction on opposite boundaries, while states with opposite spins propagate in opposite directions along a given boundary as required by time-reversal symmetry. Recent studies have demonstrated antihelical edge states in photonic systems through engineered coupling phases. Xie et al. introduced direction-dependent complex couplings in a coupled-resonator lattice to realize a photonic topological metal with antihelical edge states [31]. Liu et al. exploited pseudo-spin-dependent winding phases associated with higher-order resonator modes to achieve analogous boundary transport [32]. In contrast, a natural yet largely unexploited strategy in photonics is to introduce staggered next-nearest-neighbor (NNN) couplings within a conventional Kane-Mele framework [27, 33, 34]. This staggered NNN coupling produces spin-dependent energy shifts of the Dirac cones and thereby provides the underlying mechanism for antihelical edge states.

In this work, we establish antihelical edge transport within a modified Kane-Mele framework implemented in electromagnetic-dual photonic metacrystals. Reversing the spin-dependent next-nearest-neighbor coupling on one sublattice reorganizes the bulk spectrum into a spin-polarized Dirac semimetal (SP-DSM) characterized by pseudospin-dependent Dirac-cone energy shifts. In zigzag ribbon geometries, this bulk Dirac-cone reorganization produces tilted dispersions and gapless antihelical edge transport coexisting with bulk modes. In finite geometries, the symmetry-dependent matching of these antihelical boundary modes governs their behavior at structural corners, giving rise to corner-localized modes at 120° corners. To assess practical feasibility, we propose a realistic metacrystal based on split-ring metaatoms with opposite gap orientations to implement the required staggered bianisotropy. Furthermore, exploiting the tensorial freedom of bianisotropic constitutive relations, we show that the electromagnetic pseudospin basis can be continuously manipulated and spatially reconfigured while preserving the boundary-localized transport.

***Tight-binding Models and Antihelical Edge States. ---*** We begin with the conventional Kane-

Mele model on a honeycomb lattice, shown in Fig. 1(a). In this model, intrinsic SOC is introduced through complex NNN hopping processes with opposite circulation directions for the two spin sectors. Spin-up and spin-down sectors are indicated in blue and red, respectively, and the solid and hollow arrows denote clockwise and counterclockwise NNN hopping orientations. The introduction of this uniform SOC opens a bulk gap and gives rise to helical boundary states. Figure 1(b) presents the projected band structure of a zigzag ribbon, in which a pair of boundary modes traverse the gap. The boundary-state dispersions appear in purple, indicating the coincidence of the spin-up and spin-down sectors in the projected spectrum. The resulting boundary transport configuration is illustrated schematically in Fig. 1(c). Along a given boundary, states with opposite spins propagate in opposite directions, with their propagation direction locked to the spin degree of freedom. Reversing the boundary orientation reverses the propagation direction while preserving the spin-momentum locking, yielding the conventional helical configuration of the QSH phase.

To access a qualitatively different regime, we employ a modified Kane-Mele model featuring staggered intrinsic spin-orbit coupling [28], in which the spin-dependent NNN hopping coefficients are reversed on one of the two sublattices, as shown in Fig. 1(d). For spin-up states, the NNN hopping on the A sublattice is counterclockwise, whereas that on the B sublattice is clockwise. The hopping directions are reversed for spin-down modes. As a result, the originally uniform SOC becomes staggered between the two sublattices. The tight-binding Hamiltonian is written as

$$H = t\sum_{\langle i,j\rangle,\alpha} c_{i\alpha}^{\dagger} c_{j\alpha} + i\lambda_{\mathrm{SO}} \sum_{\xi=A,B} \eta_{\xi} \sum_{\langle\langle i,j\rangle\rangle_{\xi},\alpha\beta} \nu_{ij}\, c_{i\alpha}^{\dagger} (s_z)_{\alpha\beta} c_{j\beta}, \tag{1}$$

where $c_{i\alpha}^{\dagger}$ $(c_{i\alpha})$ creates (annihilates) an excitation with pseudospin $\alpha \in \{\uparrow, \downarrow\}$ at site i, $t$ is the nearest-neighbor hopping amplitude and $\lambda_{\mathrm{SO}}$ is the strength of NNN SOC, $s_z$ is the Pauli matrix in pseudospin space, and $\langle\langle i,j\rangle\rangle_{\xi}$ denotes NNN hopping within sublattice $\xi = A, B$. Here, $\nu_{ij} = +1$ (-

1) if the NNN hopping from site j to site i is counterclockwise (clockwise) with respect to the center of the hexagon, and $\eta_\xi = \pm 1$ is a sublattice-dependent factor describing whether the NNN SOC term on sublattice $\xi$ is retained or reversed. In the conventional model, $\eta_A = \eta_B$, the opposite orientations of the NNN hopping paths on the two sublattices, leads to effective NNN SOC contributions with opposite signs. In the modified model, reversing the NNN SOC term on one sublattice corresponds to $\eta_A = -\eta_B$, making the effective contributions identical. Because the NNN hopping acts within each sublattice, its low-energy contribution can be decomposed as

$$H_{\mathrm{NNN}} = \mathrm{diag}(\varepsilon_A, \varepsilon_B) \otimes s_z = \left[\frac{\varepsilon_A+\varepsilon_B}{2}\sigma_0 + \frac{\varepsilon_A-\varepsilon_B}{2}\sigma_z\right] \otimes s_z, \tag{2}$$

where $\varepsilon_A$ and $\varepsilon_B$ represent the effective NNN diagonal coupling strengths on sublattices A and B, respectively. $s_z$ is the operator in the spin space. In the conventional Kane–Mele model, $\varepsilon_A = -\varepsilon_B$, so the NNN interaction produces a $\sigma_z \otimes s_z$ Dirac mass and opens a bulk gap. In the modified model considered here, reversing the NNN coupling on one sublattice gives $\varepsilon_A = \varepsilon_B$. The mass term therefore vanishes, leaving only a sublattice-independent contribution proportional to $\sigma_0 \otimes s_z$. Expanding around the K and K' valleys gives the low-energy Hamiltonian

$$H_{\mathrm{eff}}(\delta\mathbf{k}) = v_D\left(\tau_z\sigma_x\delta k_x + \sigma_y\delta k_y\right) + \lambda\tau_z s_z\sigma_0, \tag{3}$$

where $v_D$ is the Dirac velocity and $\lambda$ denotes the effective SOC strength. $\tau_z$, and $\sigma_i$ are operators in the valley and sublattice spaces, respectively. The second term shifts the two pseudospin-resolved Dirac cones in opposite directions in frequency but, being proportional to $\sigma_0$, does not generate a Dirac mass.

The band structure of a zigzag ribbon of this modified model is displayed in Fig. 1(e). Instead of opening a bulk gap, the staggered SOC produces an energy displacement of the bulk Dirac cones. Specifically, the spin-up Dirac cone is shifted to higher energy at the K' valley relative to the K valley, while the opposite shift occurs for spin-down modes. These bulk Dirac-cone shifts are reflected in the projected spectrum, leading to spin-dependent tilted edge-state dispersions. Along a given boundary, spin-up and spin-down states counterpropagate as required by time-reversal symmetry [Fig. 1(f)]. However, in contrast to the helical case shown in Fig. 1(c), the spin-momentum locking is inverted between the upper and lower boundaries, such that states of a given spin propagate in the same direction along both edges. This configuration defines the antihelical edge state. To further confirm this antihelical behavior, Fig. 1(g) shows excitation results for the modified model. At the central frequency of the edge states, boundary-localized transport is observed for both spin states. Simultaneously, owing to the gapless nature of the bulk spectrum, bulk transport coexists at this exact frequency. These excitation results visualize the transport characteristics of the antihelical edge states supported by the modified model.

***Electromagnetic-Dual Photonic Crystal and Bulk Dirac Cone Reorganization. ---*** To implement the above modified model within a realistic photonic platform, we construct an electromagnetic-dual photonic crystal composed of bianisotropic rods arranged in a honeycomb lattice [Fig. 2(a)]. The constituent rods obey the constitutive relations $\boldsymbol{D} = \hat{\varepsilon} \cdot \boldsymbol{E} + i\hat{\chi} \cdot \boldsymbol{H}$ and $\boldsymbol{B} = \hat{\mu} \cdot \boldsymbol{H} - i\hat{\chi}^{\mathrm{T}} \cdot \boldsymbol{E}$. Electromagnetic duality is imposed by choosing identical relative permittivity and permeability tensors, with $\varepsilon_{xx} = \varepsilon_{yy} = \mu_{xx} = \mu_{yy} = 1$ and $\varepsilon_{zz} = \mu_{zz} = 14$. Under this condition, the TE and TM modes become degenerate, allowing the vectorial electric–magnetic polarization degree of freedom of the electromagnetic field to serve as a pseudospin degree of freedom. Specifically, the two pseudospin states are formed by the linear combinations TE ± TM, corresponding to the in-phase and out-of-phase relations between $E_z$ and $H_z$, respectively [7, 35]. The resulting pseudospin is therefore encoded directly in the electromagnetic polarization state rather than in an auxiliary structural degree of freedom. Within this electromagnetic pseudospin basis, the effective spin-orbit interaction can be generated through an antisymmetric magnetoelectric coupling tensor, parameterized by $\chi_{yx} = -\chi_{xy} = \zeta$. The magnitude and sign of $\zeta$ determine the strength and sign of the effective SOC. This provides a particularly convenient route to realize the staggered interaction required here: by assigning opposite magnetoelectric couplings to the two sublattices, $\zeta_1 = -\zeta_2$, a sublattice-staggered bianisotropy is obtained directly. The resulting configuration therefore serves as a simplified electromagnetic realization of the modified Kane–Mele model, while a realistic metacrystal implementation is presented below.

When the bianisotropic coupling parameter is identical on the two sublattices ($\zeta^{(1)} = \zeta^{(2)} = 0.9$), the system exhibits a uniform effective SOC, which introduces a mass term that lifts the Dirac-point degeneracy, opening a full band gap characteristic of the QSH phase [Fig. 2(b)]. At each valley, the original fourfold degeneracy is lifted: the twofold-degenerate modes of a given pseudospin split apart in frequency, distinguished by their distinct orbital angular momentum, with the opposite pseudospin exhibiting the reversed splitting pattern. The eigenfield profiles at the K point [Fig. 2(d)] show that the resulting four modes retain dipolar field patterns inherent to the effective medium and can be identified by their pseudospins and circulating Poynting-flux patterns.

In contrast, when the magnetoelectric coupling parameters take opposite signs on the two sublattices ($\zeta^{(1)} = 0.9, \zeta^{(2)} = -0.9$), the effective SOC becomes staggered. The Dirac cones remain gapless but are displaced in energy in a pseudospin-dependent manner [Fig. 2(c)], realizing a SP-DSM phase that reproduces the bulk structure of the modified Kane-Mele model introduced in Fig. 1(d). In this regime, as shown in Fig. 2(f), modes with opposite pseudospins are separated in frequency, while modes with the same pseudospin remain degenerate at the Dirac point. The continuous evolution of the four low-energy modes at K point between these two regimes is illustrated in Fig. 2(e), where $\zeta^{(2)}$ is tuned continuously from 0.9 to -0.9 while $\zeta^{(1)}$ is fixed at 0.9. Throughout this evolution, the pseudospin and orbital angular momentum remain well-defined quantum numbers that label each mode, allowing their identities to be tracked from the uniform regime to the staggered regime. When $\zeta^{(2)}$ reaches the opposite value of $\zeta^{(1)}$, the bulk gap closes at a critical point and the Dirac cones reorganize into the gapless configuration, defining the phase boundary between the QSH phase and the SP-DSM. The preservation of the pseudospin degree of freedom throughout this evolution demonstrates the correspondence between our electromagnetic-dual platform and the modified Kane–Mele model. Notably, the spin-polarized Dirac crossings of the SP-DSM phase are maintained only when the two pseudospin channels remain decoupled under electromagnetic duality ($\hat{\varepsilon} = \hat{\mu}$) and the staggered magnetoelectric couplings remain balanced ($\zeta^{(1)} = -\zeta^{(2)}$). Breaking duality hybridizes the TE±TM channels, while an imbalance $\zeta^{(1)} \neq -\zeta^{(2)}$ generates an effective mass term and opens a Dirac gap.

***Antihelical Edge Transport and localized Corner Mode. ---*** To examine the boundary manifestations under the staggered effective SOC, we construct a ribbon geometry periodic along the x direction with zigzag terminations [Fig. 3(a)]. The sublattices are assigned opposite magnetoelectric coupling parameters, $\zeta^{(1)} = 0.9$ and $\zeta^{(2)} = -0.9$, consistent with the SP-DSM phase studied before. This boundary configuration preserves valley distinguishability, projecting K and K' valleys onto distinct momenta $k_x$. The projected band structures of the ribbon for the two pseudospin sectors are shown in Figs. 3(b) and 3(c). Since the bulk Dirac cones remain gapless while being displaced in frequency, the projected bands exhibit a characteristic tilt in both cases within the momentum interval connecting the projected Dirac points (i.e., $k_x \in [2\pi/3a, 4\pi/3a]$). This tilt leads to directional edge dispersions, with the branches associated with the upper and lower zigzag boundaries sharing the same sign of slope. Specifically, the spin-up sector exhibits a positive slope [Fig. 3(b)], whereas the spin-down sector shows a negative slope [Fig. 3(c)], giving rise to an antihelical edge-state configuration. The staggered magnetoelectric coupling coefficients induces a lattice-dependent effective spin-orbit interaction, which shifts the Dirac cones in a pseudospin-dependent manner and gives rise to the observed tilt of the projected bands.

Due to the gapless nature of the spectrum, the edge branches inevitably overlap spectrally with bulk states. As illustrated in Figs. 3(b) and 3(c), a horizontal line at the center frequency of the edge-state dispersion intersects the projected bands at four momenta: two located within the interval $k_x \in [2\pi/3a, 4\pi/3a]$, corresponding to edge modes, and two outside this interval, corresponding to bulk-like modes. This leads to the coexistence of boundary-localized and extended states at the same frequency. The spatial field profiles at this frequency are shown in Figs. 3(d) and 3(e), where the pseudospin-resolved field amplitudes are defined as $|\mathbf{\Psi}^{\uparrow}| = |\sqrt{\varepsilon_0}E_z + \sqrt{\mu_0}H_z|$ and $|\mathbf{\Psi}^{\downarrow}| = |\sqrt{\varepsilon_0}E_z - \sqrt{\mu_0}H_z|$. For momenta within the tilted dispersion range, the fields are strongly localized near the ribbon boundaries (panels 2 and 3). In contrast, for momenta outside this interval, the modes extend across the ribbon (panels 1 and 4). Although the edge and bulk-like modes coexist at the same frequency, their distinct momenta and transverse localization suppress direct coupling at a translationally invariant boundary. Perturbations that provide sufficient momentum transfer, however, can induce appreciable scattering into bulk-like states. Transport simulations further corroborate the antihelical nature of these modes. As shown in the left panels of Figs. 3(f) and 3(g), edge excitations at the center frequency yield explicitly directional boundary transport: spin-up sources excite right-propagating waves, whereas spin-down sources result in left-propagating waves. Interestingly, for a given pseudospin, placing the excitation source in the interior of the ribbon couples to bulk-like modes that propagate in the opposite direction relative to their corresponding boundary modes [right panels of Figs. 3(f) and 3(g)].

The directional nature of the antihelical boundary states also governs their behavior at geometric corners. In the present staggered-bianisotropy configuration, $C_3$ symmetry is strictly preserved, whereas $C_6$ symmetry is broken. As a consequence, geometrically distinct corners are not equivalent from the viewpoint of boundary-mode matching. For a 60° interior corner, the two adjoining edges are related by a $C_3$ operation, so that their boundary modes possess compatible symmetry and propagation characteristics. The antihelical mode can therefore connect continuously from one edge to the other and propagate through the corner without forming a localized state. By contrast, the two adjoining edges of a 120° corner are not related by the preserved $C_3$ symmetry. The corresponding boundary modes are therefore mismatched in their symmetry character and cannot be smoothly connected across the corner, allowing their coupling to produce a localized corner eigenmode. To explicitly examine this finite-geometry effect, we construct photonic-crystal flakes with a parallelogram geometry. The flake contains two 120° corners and two 60° corners [Fig. 4(a)], providing both types of terminations within the same finite structure. The corresponding eigenfrequency spectrum [Fig. 4(b)] exhibits discrete finite-size modes associated with the corners. Their spatial field distributions [Fig. 4(c)] show pronounced localization only at the two 120° corners, whereas no comparable localization appears at the 60° corners. These results demonstrate that the corner localization is determined by the symmetry-dependent matching of the antihelical boundary modes on the adjoining edges. The finite-geometry response therefore provides an additional manifestation of the directional boundary physics beyond straight-edge antihelical transport.

***Realistic Implementation. ---*** To demonstrate that the staggered bianisotropy underlying the antihelical edge states can be implemented in a realistic photonic structure, we construct a metacrystal composed of electromagnetic-dual metaatoms. The constituent metaatoms of the metacrystal are shown in Fig. 5(a). Each is designed to satisfy the electromagnetic duality condition, with balanced electric and magnetic responses over the frequency range of interest [35]. In particular, the broken mirror symmetry of the split-ring metaatoms induces a finite magnetoelectric coupling, whose sign is determined by the orientation of the structural asymmetry. Reversing the gap orientation of the split-ring resonator therefore reverses the effective bianisotropic response, providing a direct geometrical route to realize opposite bianisotropy on different sublattices. The present design operates in the microwave regime, with a Dirac-point frequency of about 5 GHz and resonator dimensions on the centimeter scale. Similar SRR-based bianisotropic structures have been experimentally fabricated and characterized in the microwave regime [36, 37].

Based on these metaatoms, a unit cell of the metacrystal is assembled, as illustrated in Fig. 5(b). In particular, two split-ring metaatoms with opposite orientations are placed on the two sublattice sites of the honeycomb unit cell, thereby generating a staggered bianisotropy. The corresponding bulk band structure is shown in Fig. 5(c). The metacrystal exhibits gapless Dirac cones located at the K and K′ valleys, with a clear pseudospin-dependent frequency displacement. This band reorganization reproduces the key features obtained from the effective-medium model in Fig. 2(c), confirming that the staggered bianisotropic coupling implemented by the split-ring orientation directly maps to the staggered effective spin-orbit interaction in the low-energy Dirac description. To further verify the agreement between the realistic metacrystal and the effective theory, the eigenfield profiles at the K and K′ points are shown in Fig. 5(d). The relative phase between the out-of-plane electric and magnetic field components clearly distinguishes the two pseudospin states through their in-phase and out-of-phase relations. The field profiles confirm that the low-frequency modes retain the same pseudospin character as those obtained in the effective-medium model. These results show that the staggered bianisotropy implemented by the split-ring metaatoms faithfully captures the Dirac-cone reorganization responsible for the antihelical edge states.

***Continuous Pseudospin Engineering via Tensorial Bianisotropy.*** **---** The constitutive relation employed above represents a particular form of bianisotropic coupling. To explore additional electromagnetic degrees of freedom that are not captured by the effective tight-binding Hamiltonian, we extend it to a general bianisotropic form, $\boldsymbol{D} = \hat{\varepsilon} \cdot \boldsymbol{E} + \hat{\chi} \cdot \boldsymbol{H}$ and $\boldsymbol{B} = \hat{\mu} \cdot \boldsymbol{H} + \hat{\xi} \cdot \boldsymbol{E}$, where $\hat{\chi}$ and $\hat{\xi}$ denote the two magnetoelectric coupling tensors. The constitutive relation adopted in the preceding sections corresponds to a particular choice of these tensors. The more general form allows their real and imaginary off-diagonal components to be continuously varied while preserving electromagnetic duality and the effective staggered SOC. We parameterize this additional tensorial degree of freedom by an angle $\theta$, with the nonzero off-diagonal components written as $\chi_{xy} = -\chi_{yx} = \zeta(-i\cos\theta - \sin\theta)$, $\xi_{xy} = -\xi_{yx} = \zeta(-i\cos\theta + \sin\theta)$. Here, $\zeta$ determines the strength of the effective spin-dependent coupling, whereas $\theta$ controls its tensorial realization. At $\theta = 0$, the coupling reduces to the response employed in the preceding sections, while at $\theta = \pi/2$ it becomes Tellegen-like. Importantly, varying $\theta$ changes the relative real and imaginary parts of the magnetoelectric coupling without changing the magnitude or staggered configuration of the effective SOC. The corresponding electromagnetic pseudospin basis evolves continuously as TE $\pm e^{i\theta}$ TM. Thus, $\theta$ simultaneously parameterizes the magnetoelectric tensor and the relative phase between the TE and TM components of the pseudospin. For example, $\theta = 0$ gives the linear combinations TE ± TM, whereas $\theta = \pi/2$ gives TE ± $i$TM. As $\theta$ varies continuously from 0 to π, the two orthogonal pseudospin states trace continuous trajectories in the $S_2$–$S_3$ plane of the Poincaré sphere [Fig. 6(a)]. The same effective antihelical Hamiltonian therefore admits a continuous family of electromagnetic pseudospin representations.

This constitutive degree of freedom can further be made spatially dependent. We consider a ribbon in which $\theta(x)$ varies smoothly from 0 to π along the propagation direction [Fig. 6(b)]. Full-wave simulations show that the antihelical mode remains localized at the boundary throughout the graded region [Fig. 6(c)], while its local electromagnetic composition evolves continuously. This evolution is directly visualized by the spatial distribution of the relative phase $\arg(E_z/H_z)$ in Fig. 6(d), which follows the prescribed variation of $\theta(x)$. The tensorial bianisotropy therefore provides an additional electromagnetic degree of freedom that allows the pseudospin representation to be continuously and spatially manipulated while preserving the underlying antihelical boundary transport.

***Conclusion.*** --- We have proposed and demonstrated a physical route to gapless antihelical edge states within a modified Kane-Mele framework realized in electromagnetic-dual photonic metacrystals. By introducing staggered bianisotropy, the bulk Dirac spectrum is reorganized into a spin-polarized Dirac semimetal characterized by pseudospin-dependent energy offsets. In zigzag ribbon geometries, this bulk reorganization produces tilted edge dispersions where boundary-localized modes coexist with extended bulk states and display antihelical transport: opposite pseudospins counterpropagate along a given boundary, while identical pseudospins propagate in the same direction on opposite edges.

Beyond straight boundaries, the symmetry-dependent matching of these boundary modes leads to corner localization specifically at 120° corners. To demonstrate practical feasibility, a realistic metacrystal based on split-ring metaatoms is designed to implement the required staggered bianisotropy in the microwave regime. Furthermore, exploiting the tensorial freedom of the constitutive relations enables continuous manipulation and spatial reconfiguration of the electromagnetic pseudospin basis while preserving antihelical boundary transport. The framework presented here bridges tight-binding models, effective-medium theory, and realistic metacrystal implementations, and provides a general strategy for exploring unconventional boundary phenomena beyond the conventional gapped topological paradigm.

**Acknowledgements**

This work was supported by National Natural Science Foundation of China (12522415, 12374364), Guangdong Basic and Applied Basic Research Foundation (2023B1515040023), Guangzhou Science, Technology and Innovation Commission (2024A04J6333).

## Figures and captions

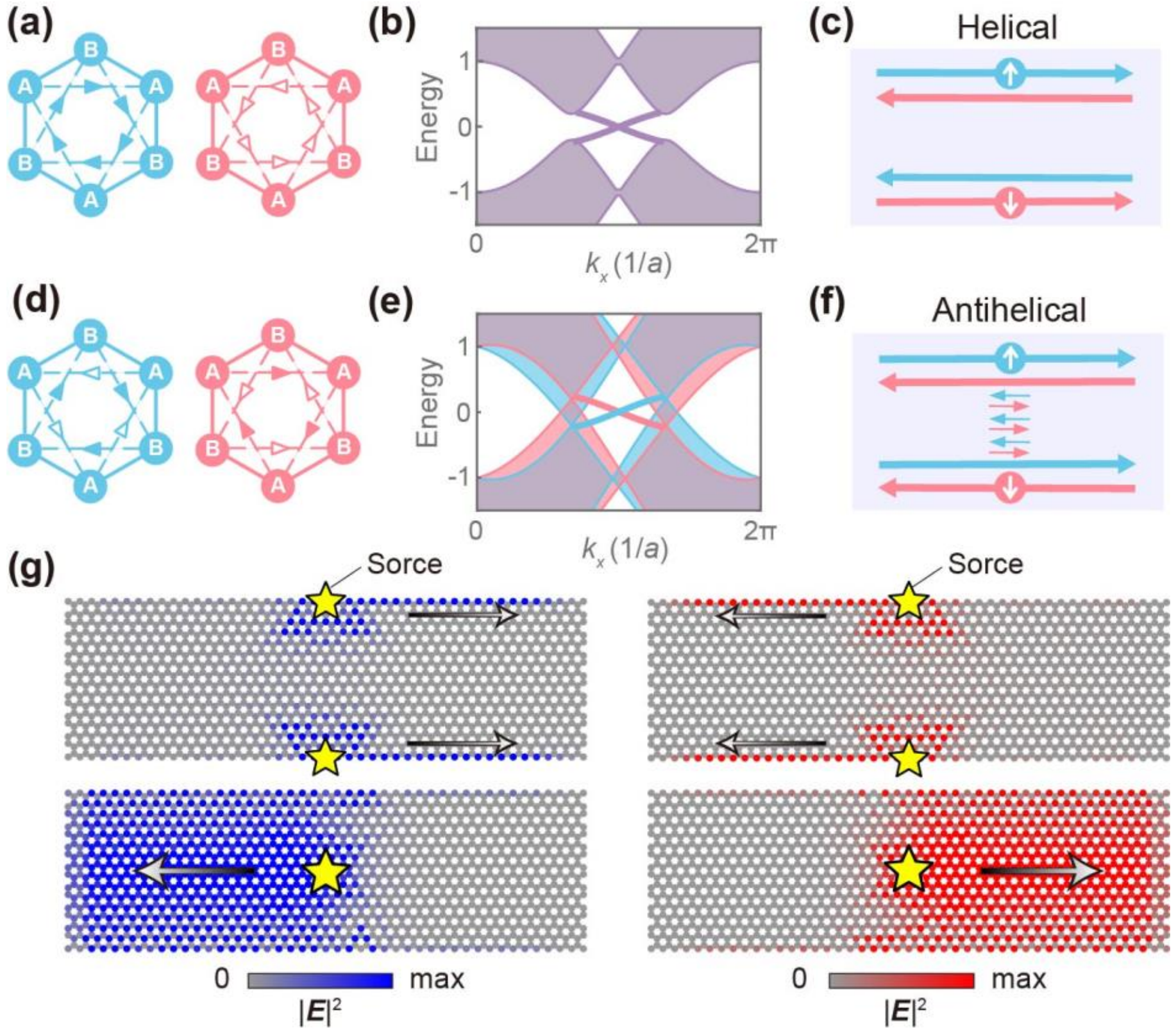

**FIG. 1. Modified Kane-Mele model and antihelical edge states.** (**a**) Conventional Kane-Mele model on a honeycomb lattice. Spin-up (blue) and spin-down (red) sectors feature uniform NNN hopping. Solid and hollow arrows denote clockwise and counterclockwise orientations, respectively. (**b**) Projected band structure of the model in (a) for a zigzag ribbon. Purple indicates the spectral coincidence of the two spin sectors, representing both bulk continua and discrete edge modes. (**c**) Schematic of helical edge states: for a given spin, the propagation direction is reversed between the upper and lower boundaries. (**d**) Modified Kane-Mele model with staggered NNN hopping, where the hopping direction on sublattice A is reversed relative to (a). (**e**) Projected band structure for the modified model. The staggered SOC induces a spin-dependent energy shift of the Dirac cones, resulting in tilted edge dispersions. (**f**) Schematic of antihelical edge states: for a given spin, the propagation direction is identical on both boundaries. The interior arrows represent the coexisting directional bulk-like modes. (**g**) Simulated excitation results for the modified model. Left/right panels show spin-up/down excitations. Upper panels: boundary sources excite localized edge transport. Lower panels: interior sources excite extended bulk-like transport.

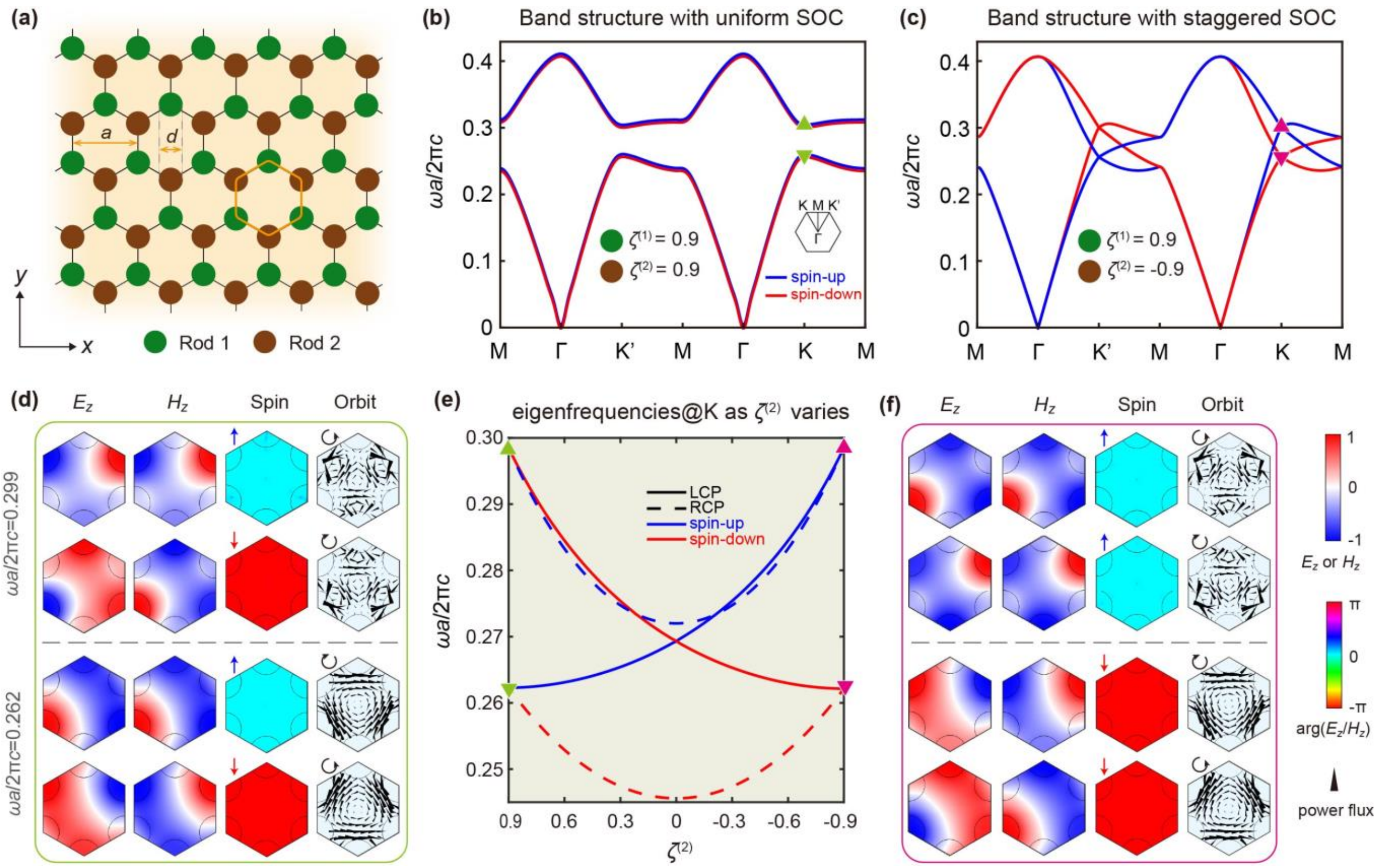

**FIG. 2. Electromagnetic-dual photonic crystal realization of the modified Kane-Mele model.** (**a**) Schematic of the electromagnetic-dual photonic crystal consisting of bianisotropic rods in a honeycomb lattice. Rod 1 (green) and Rod 2 (brown) occupy the two sublattices. Each rod is characterized by its magnetoelectric coupling $\zeta$ and satisfies the duality condition ($\hat{\varepsilon} = \hat{\mu}$). (**b**) Bulk bands in the QSH phase under uniform bianisotropic coupling ($\zeta^{(1)} = \zeta^{(2)} = 0.9$), showing a complete band gap. (**c**) Bulk bands in the SP-DSM phase under staggered bianisotropic coupling ($\zeta^{(1)} = 0.9$, $\zeta^{(2)} = -0.9$), exhibiting pseudospin-dependent energy shifts of the Dirac cones. (d) Eigenfield profiles of the four bulk modes at the K point in the QSH phase. Pseudospin (Spin) is defined by the relative phase between $E_z$ and $H_z$, while orbital angular momentum (Orbit) is inferred from the circulating power flux. (e) Evolution of eigenfrequencies of bulk modes at the K point as $\zeta^{(2)}$ varies from 0.9 to -0.9, illustrating the transition from the gapped QSH phase to the SP-DSM regime. (**f**) Eigenfield profiles of the four bulk modes at the K point in the SP-DSM phase, where modes of the same pseudospin remain degenerate while opposite pseudospins are frequency-separated.

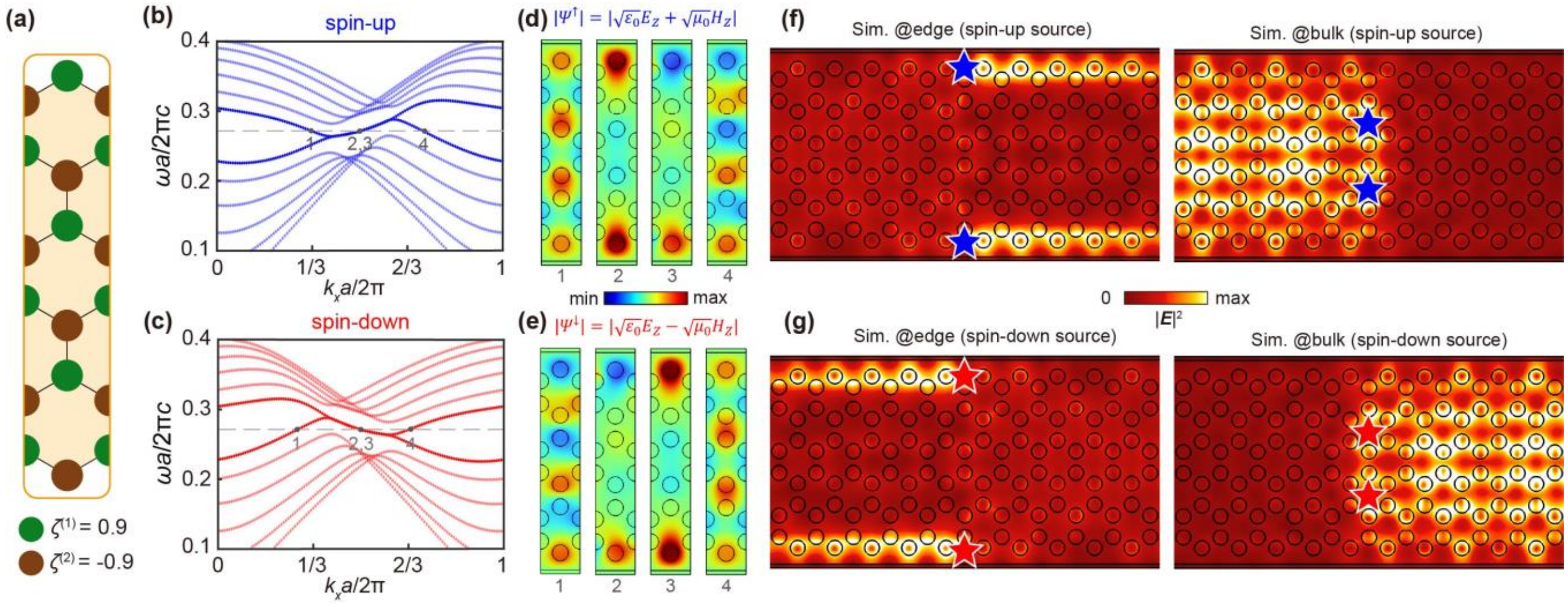

**FIG. 3. Antihelical edge states in a zigzag ribbon geometry.** (**a**) Supercell of a zigzag ribbon constructed from the electromagnetic-dual photonic crystal with staggered bianisotropy. (**b**, **c**) Projected band structures for the (b) spin-up and (c) spin-down sectors. The staggered effective SOC induces pseudospin-dependent energy shifts of the Dirac cones, yielding tilted gapless dispersions. The horizontal dashed line marks the excitation frequency for the simulations, intersecting the bands at four distinct momenta (labeled 1-4). (**d**, **e**) Spatial profiles of the pseudospin-resolved field amplitudes, $|\Psi^{\uparrow}|$ and $|\Psi^{\downarrow}|$, corresponding to the four states marked in (b) and (c). Modes 1 and 4 are extended bulk-like states, whereas modes 2 and 3 are boundary-localized edge states. (**f**, **g**) Full-wave simulations of field intensity profiles ($|\boldsymbol{E}|^2$) for (f) spin-up and (g) spin-down excitations. Left panels: Boundary sources excite directional, antihelical edge transport. Right panels: Interior sources excite directional bulk transport propagating in the opposite direction relative to the corresponding edge modes.

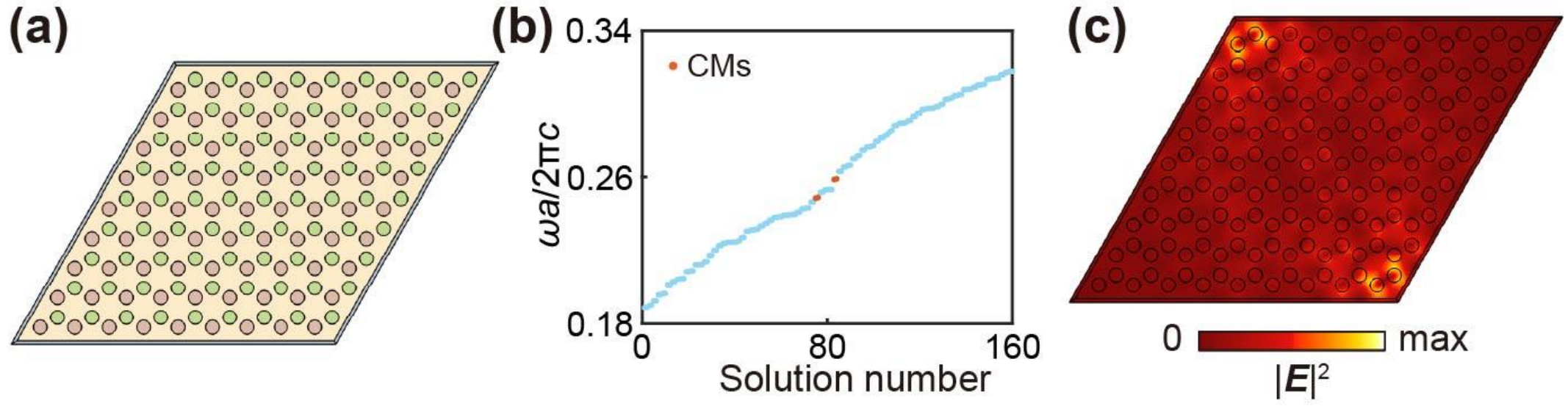

**FIG. 4. Corner-localized eigenmodes in a finite geometry.** (**a**) Schematic of a finite parallelogram photonic crystal containing two 120° corners and two 60° corners. (**b**) Eigenfrequency spectrum of the finite structure. (**c**) Spatial field intensity profile $|\boldsymbol{E}|^2$ for the corner-localized eigenmode, showing localization at the two 120° corners.

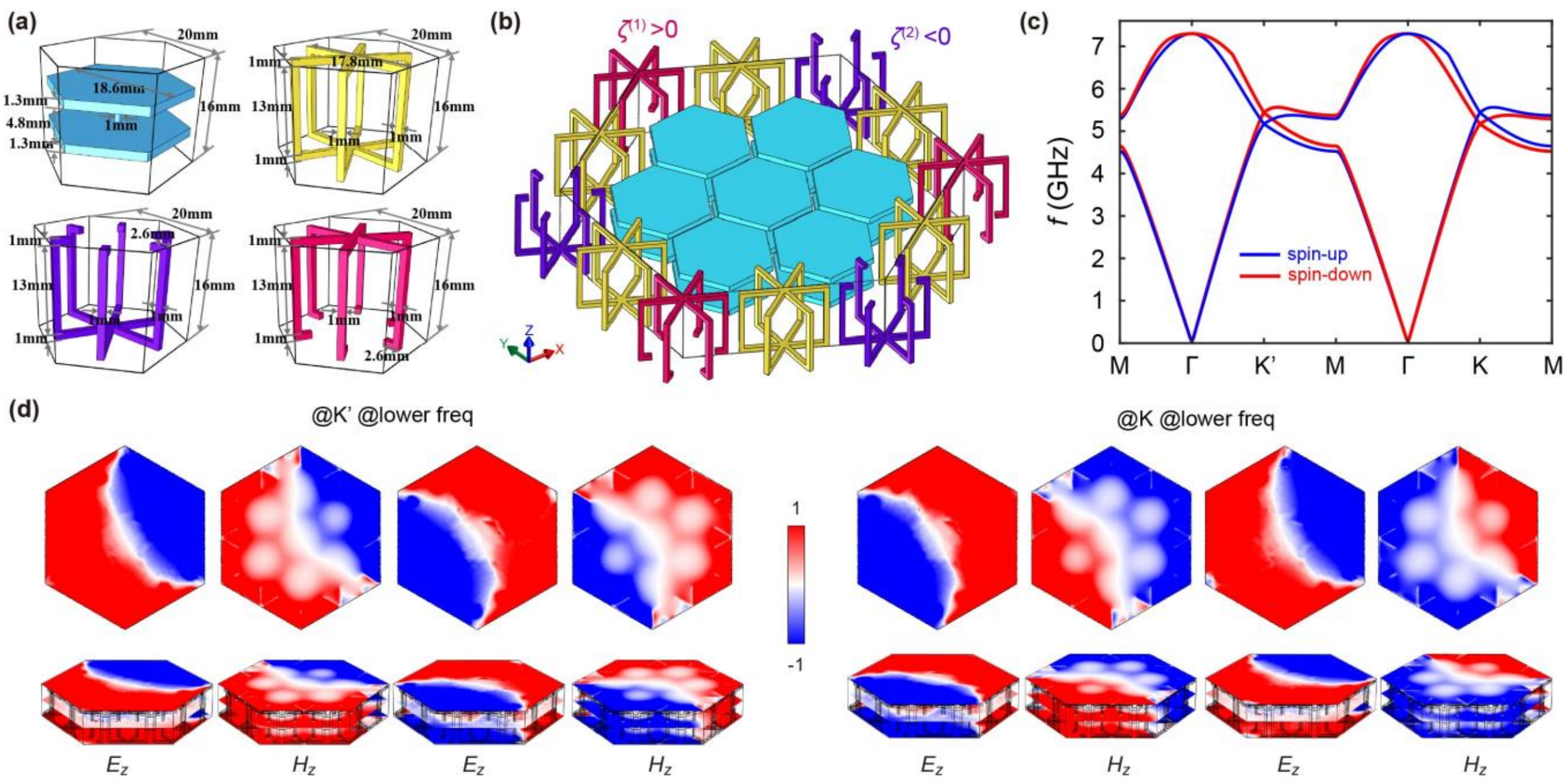

**FIG. 5. Realistic metacrystal implementation of staggered bianisotropy and spin-polarized Dirac cones.** (**a**) Design of the constituent metaatoms for the electromagnetic-dual metacrystal. The broken mirror symmetry of the split-ring resonators induces a finite magnetoelectric coupling, with the gap orientation determining the sign of the effective bianisotropic parameter. (**b**) Schematic of the assembled metacrystal. By placing split-ring metaatoms with opposite gap orientations on the two sublattices, a staggered bianisotropy is realized. (**c**) Bulk band structure of the metacrystal. Spin-up (blue) and spin-down (red) bands exhibit gapless Dirac cones at the K and K′ valleys with a pseudospin-dependent frequency displacement, faithfully reproducing the effective-medium results in Fig. 2(c). (**d**) Eigenfield profiles of the bulk modes at the K and K′ points (extracted at the lower-frequency branches). The pseudospin character is identified by the relative phase between $E_z$ and $H_z$ components.

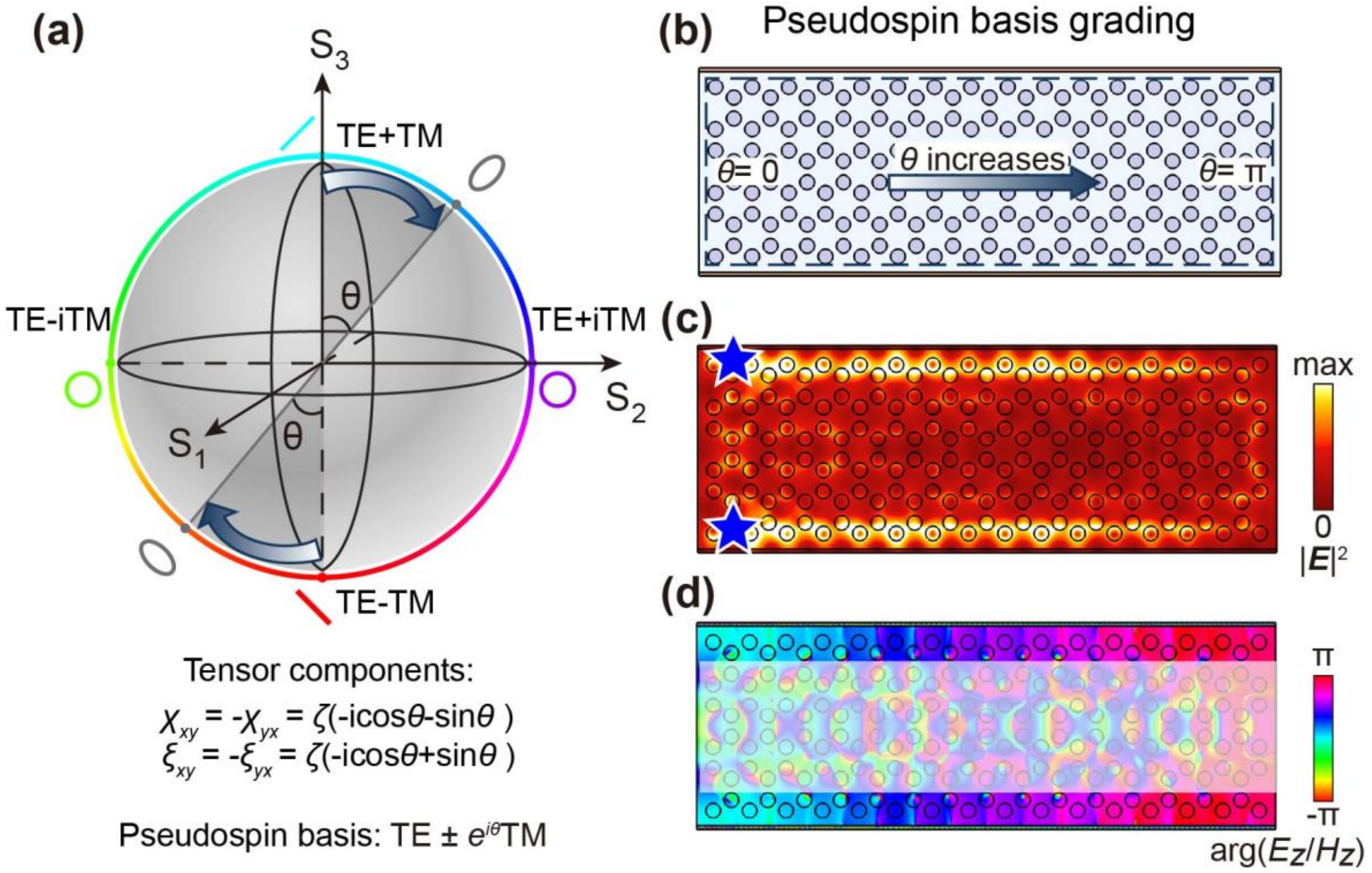

**Fig. 6. Continuous engineering of the electromagnetic pseudospin through tensorial bianisotropy.** (**a**) Evolution of the pseudospin states TE ± $e^{i\theta}$TM on the $S_2$–$S_3$ plane of the Poincaré sphere as $\theta$ varies from 0 to π. The parameter $\theta$ controls the relative real and imaginary components of the magnetoelectric tensors, while the effective staggered SOC is preserved. (**b**) Schematic of a photonic ribbon with a graded $\theta(x)$, varying smoothly from 0 to π. (**c**) Simulated electric-field intensity $|\boldsymbol{E}|^2$, showing that the antihelical mode remains localized to the boundary throughout the graded region. (**d**) Spatial distribution of arg($E_z/H_z$) showing the continuous evolution of the local electromagnetic pseudospin along the propagation direction.